\documentclass[12pt,a4paper]{article}
\pdfoutput=1 
\usepackage[latin1]{inputenc}
\usepackage[T1]{fontenc}
\usepackage{amsmath}
\usepackage{amsfonts}
\usepackage{amssymb}
\usepackage{graphicx}
\usepackage{pslatex}
\usepackage[a4paper,  margin=50pt]{geometry}
\usepackage{setspace}
\usepackage{environ}
\usepackage{nicefrac}
\usepackage{fancyhdr}

\usepackage[backend=biber, style=authoryear, uniquename=false, maxcitenames=2, mincitenames=1, natbib]{biblatex}
\NewEnviron{myequation}[1]{%
\begin{equation}
\setstretch{2}
\begin{array}{ccc}
\BODY
\end{array}
{#1}
\end{equation}%
}

\title{Earthquakes big and small: same physics, different boundary conditions}
\author{Stefan Nielsen\footnote{
		Earth Science Department, Durham University, UK. 
		stefan.nielsen@durham.ac.uk}}
\begin{document}

\everymath={\displaystyle}
%%%%%%%%%%%%%%%%%%%%%%%%%%%%%%%%%%%%%%%

\maketitle	

%\tableofcontents\newpage

\section*{Abstract}
\addcontentsline{toc}{section}{Abstract}
Self-similarity indicates that large and small earthquakes share the same physics, where all variables scale with rupture length $L$. Here I show that rupture tip acceleration during the
start of dynamic rupture (break-out phase) is also self-similar, scaling with $L_c$ in space and $L_c/C_{lim}$ in time (where $L_c$ is the breakout patch length and $C_{lim}$ the limiting rupture velocity in the subsonic regime). Rupture acceleration in the breakout phase is slower for larger initial breakout patches $L_c$. Because small faults cannot host large breakout patches, a large and slower initial breakout may be indicative of a potentially large final earthquake magnitude. Initial moment rate $\dot{M}_o$ also grows slower for larger $L_c$, therefore it may reflect fault dimensions and carry a probabilistic forecast of magnitude as suggested in some Early Warning studies. This result does not violate causality and is fully compatible with the shared fundamental, self-similar physics across all the magnitude spectrum. 

%\section*{Key Points}
%\addcontentsline{toc}{section}{Key Points}
%
%\begin{itemize}
%	\item A simple analytical solution describing the acceleration of the rupture tip in the initial rupture stages is derived
%	\item Rupture breakout from a larger nucleation patch starts slower
%	\item Larger breakout patches can be hosted only on larger fault structures
%	\item A statistical correlation between maximum rupture length and initial moment rate is expected   
%\end{itemize}

\section{Introduction}

Earthquakes are self-similar, indicating that earthquake physics is the same at all scales. 
As a direct consequence of elasticity, the rupture length dimension $L$ controls 
the scaling of slip 
($U \propto L \Delta\tau/\mu'$, where $\Delta\tau$ is the stress drop and $\mu'$ is the shear modulus), 
of duration 
($T \propto L/v_r$, where $v_r$ is the rupture velocity) 
and of energy flow 
($G \propto L\Delta\tau^2/\mu'$). 
Given the above scaling relations, and given that no systematic change of stress drop $\Delta \tau$ or rupture velocity $v_r$ with magnitude is observed, it is widely recognised that small earthquakes are identical to large ones, except for their size. 

However the starting and the stopping phases of earthquake rupture are controlled by
complex boundary conditions, dictated by fault geometry and fault segmentation where, arguably, it is less obvious to demonstrate scale independence. Inspection of exposed fault surfaces and fault traces reveals (1) self-affine, rather than self-similar topography of individual fault patches \citep{Candela2012,Bistacchi2011}, (2) segmentation of faults often creating en echelon structures of variable dimensions and separations \citep{Wesnousky2006}, and (3) difference in the roughness of large, mature faults and smaller faults with reduced cumulative offset \citep{Sagy2007}. 

Regarding the stopping phase. To stop larger earthquakes, a larger energy dissipation (critical fracture energy $G_c$) is necessary to counteract the energy flow $G$. $G_c$ is usually intended as a material property in fracture mechanics, and not related to any of the scaling relations defined above in the rupture solution. Therefore the origin of the scaling of the apparent $G_c$ with $L$ in earthquakes \citep{Abercrombie2005,Nielsen2016a} is less clear. Off-fault damage and gradual breakdown energy increase related to protracted dynamic frictional weakening have been proposed as possible origins for such scaling, as supported by field 
\citep{Shipton2006,Dor2006,Doan2009}  
numerical modelling 
(\cite{Dunham2011, Xu2012a},b)
and experimental 
(\cite{Nielsen2016a}a,b; \cite{Aben2020})  
evidence. To indicate a general parameter that includes such additional sources of energy dissipation, I will use $\Gamma$ in place of $G_c$ henceforth (where $\Gamma= G_c + \textrm{alt}$).

\begin{figure}[h]
	\centering
	\includegraphics[width=0.7\linewidth]{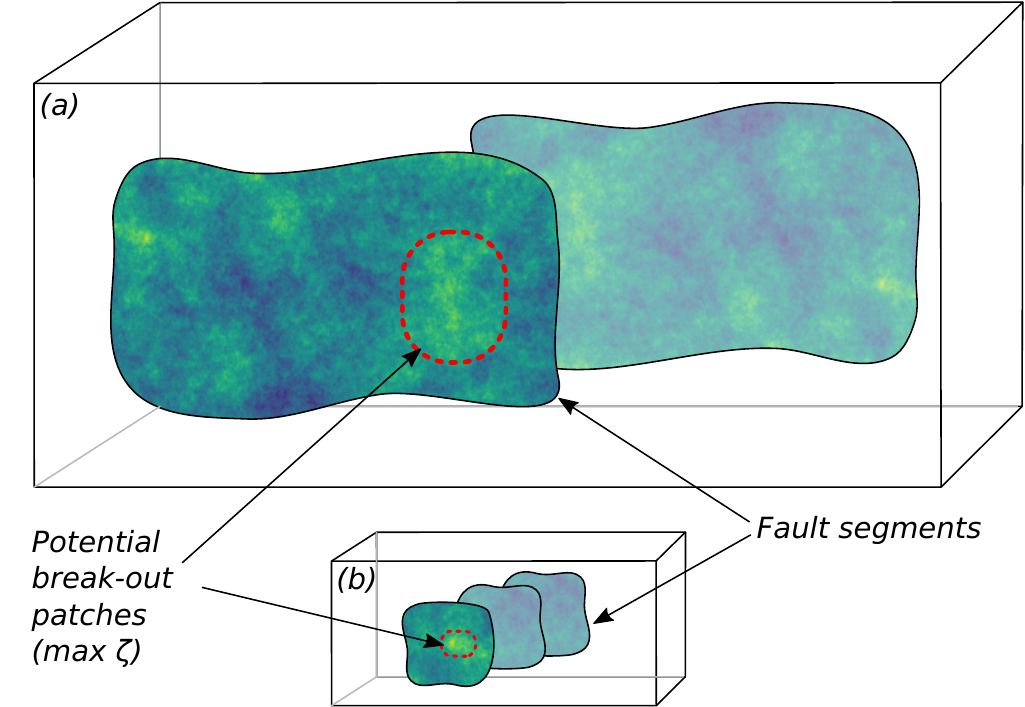}
	\caption{Conceptual illustration of rupture break-out patches 
		on fault segments of different sizes. 
		The most likely breakout patch is the area that maximizes $\zeta$  (equation \ref{eq:minstress}), where
		$\zeta$ scales with the area dimension. 
		In any case, the maximum possible break-out dimension is limited by the fault segment size.  (a) case of a mature fault with large segments, 
		and a higher probability of yielding a large magnitude 
		earthquake; (b) immature fault with smaller segments, less likely to generate large magnitude earthquake. A larger break-out patch will initially expand more slowly.
	}
	\label{fig:concept}
\end{figure}

If, on average, small earthquakes are identical to large earthquakes and take place on identical faults, why does rupture propagation stops early in most cases, and occasionally continues to propagate to larger extents with a progressively lower probability? This may be the consequence of any of the following processes:

(1) Earthquake rupture develops in a system close to critical state,
maintained near its energy threshold by self-organisation (self-organised criticality \citep{Bak1989}. 
According to this model, the earthquake process is analogous to a pile of sand close to the instability slope, where avalanches of all sizes are generated. 

(2) Dynamically generated, off-fault ductile strain increases proportionally to $L$, dissipating more energy to the detriment of rupture propagation, so that a growing rupture will carry its self-demise. Similarly, it can be argued that large earthquakes are hosted by large faults that have additional built-in dissipative power (e.g., a wider damage zone inherited from the cumulative slip history). In such case, large and small earthquakes may be similar qualitatively but not quantitatively.

(3) There exist upper limits to rupture area, dictated by natural boundary condition. These are determined by a number of structural features such as the thickness of the brittle crust, the depth of the subducting thrust, or the presence of velocity-strengthening patches that can sustain slow slip or creep but do not allow fast slip associated with dynamic earthquake rupture. 

(4) A more subtle boundary condition is determined by fault segmentation at all scales. This is a lesser constraint because rupture may propagate through minor geometrical features such as bends, jogs, or even across steps across two separate nearby segments under the favourable circumstances as shown in numerical tests \citep{Harris1993,Harris1999,Harris2002}. Therefore such geometrical segmentation constitutes a weaker boundary constraint that may be considered from a probabilistic perspective. In 22 strike-slip earthquakes where fault traces were visible at the surface, \citet{Wesnousky2006} investigated how often fault steps
({\em en echelon} features) act as barriers to rupture, among a total of 52 steps and 10 apparent fault terminations. He found that all ruptures failed to propagate across steps larger than 4 km (distance measured perpendicular to the average fault trace) and across terminations. Below 4 km, the ratio of steps where ruptures stopped to those where it continued was 17:25 (excluding terminations), with decreasing number of ruptures propagating as segment separation increased. A rough estimate from these observations sets the overall fraction of ruptures that cross any step in this dataset to $\approx$32\%. These statistics are biased by the nature and number of  events: a modest number of suitable earthquakes were used are all of which are strike-slip and reaching the surface. However these observations clearly indicate that major steps exist and that they do act as a boundary condition to rupture propagation more often than not.

Regarding the starting phase. I would like to mark a distinction between the quasi-static nucleation under slow slip and the proper break-out phase of the earthquake with rupture propagation at seismic or sub-seismic velocity. Rate-and-state friction is an appropriate model for slow-slip dynamics during the transient, possibly protracted instability phase preceding an earthquake. The early nucleation process does not necessarily imply propagation of rupture, but may show instability and slip acceleration within a delimited area of the fault that does not necessarily grow (see modelling results in \cite{Dieterich1992, Rubin2005}). Although I briefly discuss below some of the similarities between instability from the rate-and-state perspective and rupture propagation from the energy and fracture mechanics viewpoint, my focus here is rather on the break-out phase where rupture is propagating at a sensitive fraction of elastic waves velocity.

In principle, both small and large earthquakes may start from an equally small breakout patch; the final extent of the rupture would be determined through processes (1-4) outlined above. A small initial breakout may cascade across fault segments of increasing sizes \citep{Ellsworth1995} to generate a large magnitude earthquake. However, the cascade model does not consider the role of segment size as a boundary condition during rupture nucleation and breakout phases. At the other end of the spectrum, the pre-slip model (ibid) assumes that the breakout of dynamic rupture initiates from an extended patch of slow slip, thus implicitly considering the boundary conditions imposed by fault segmentation. 

The cascade versus pre-slip model may oversimplify the complex nature of 
earthquake initiation \citep{MartinezGarzon2024}, 
where the actual behaviour results from a combination of processes and is to date not fully understood. Keeping this in mind, rupture initiation documented on inter-plate and intra-plate earthquakes shows that earthquakes in the former category are preceded by measurable slow-slip episodes \citep{Bouchon2013,Kato2012,Yao2020,Bouchon2022}, while the latter do not; this suggests that intra-plate earthquakes may be closer to the cascade model while inter-plate ones may reflect a pre-slip process. Although this is highly debatable, one possible source of such difference is the increased segmentation of intra-plate faults that have a lesser degree of maturity than inter-plate ones. 

Here, I show that rupture acceleration during the breakout process scales with the inverse length of the initial rupture patch $L_c$ where $\dot{v}_r\propto 1/L_c$, a result that I derive from classic fracture mechanics regarding unstable propagation of rupture. I then derive an approximate analytical expression for the rupture front position $L(t)$ and show that it exhibits self-similarity in $L/L_c$.  This result is also mirrored by laboratory experiments and numerical simulations, that demonstrate an equivalent self-similar behaviour also in 3D.

In addition, I discuss nucleation and breakout on segments of different dimensions $L_f$. Because $L_c \leq L_f$, larger fault segments can host larger critical breakout patches, and  will show larger values of $L_c$ on average, if $L_c$ is drawn randomly from a distribution with a higher upper bound (Figure \ref{fig:concept}). I conclude that if the segment size $L_f$ controls to some extent the upper limit magnitude of the earthquake, and also the expected $L_c$ value, then large earthquakes will have on average larger breakout patches and a slower initial rupture acceleration. 
As a consequence, I suggest that the scaling relation $\dot{v}_{r} \propto 1/L_f$ in the breakout phase may also apply, in particular on inter-plate faults where rupture initiates according to the pre-slip model.

To exemplify a possible statistical relation between $L_c$ and $L_f$, I assume a self-affine distribution of parameters relevant for rupture on the fault, and compute the fracture energy for different sub-fault segments sizes $L < L_f$, finding the size $L_c$ closest to rupture in a number of random realizations. For the specific parameter distribution used here, I find that on average $L_c \approx 0.23 \times  L_f$, and estimate that $L_c \geq 0.25 \times L_f$ in approximately 63\% of the cases, with no case of $L_c < 0.125\times L_f$ among 30 realisations.

Finally, I discuss these findings in terms of far-field radiation and the possible implications for Early Warning  studies. Particle displacement in the far-field radiation is proportional to the moment rate \citep{Aki2002}.
The initial moment rate of an earthquake will depend on the area of the break-out and on its early rate of growth. According to the rupture front propagation solution proposed here, 
the initial moment rate grows relatively faster for a small initial breakout area, which indicates a statistically smaller host fault segment, and therefore a reduced probability of a large magnitude earthquake. Conversely, a slower onset of far-field motion is indicative of a slower moment rate growth and an extended break-out patch on a large fault segment, therfore, a larger probability that the fault segment may host a large magnitude earthquake. Based on analysis of earthquake data, it has been argued 
\citep{Ishihara1992,Colombelli2014,Colombelli2020} 
that the initial seconds of a seismogram may be indicative of the earthquake final magnitude, although this is matter of controversy 
\citep{Meier2017} and, so far, was lacking the theoretical support of a proper causative model. In addition, it should be noted that the initial seconds of a seismogram may include effects from later phases of rupture, due to the causal kinematics of radiation from an extended source, which entails directivity and relative position of source and receiver \citep{Murphy2009}. However these effects can be accounted for and corrected; therefore, the implications of fault segmentation and initial rupture growth model can in principle be tested on a suitable database of seismological observations.

\section{Analytic solution for rupture tip acceleration}

The crack tip motion problem can  be formulated \citep{Freund1990} as an equation where rupture velocity-dependent energy flow $G$ per unit advancement of the rupture should match $\Gamma$ (the energy dissipated in the breakdown process):
\begin{equation}
\Gamma = G(\gamma)
\label{eq:eflow1}
\end{equation}
where, in 2D sub-sonic cases, $\gamma=v_r/C_{lim}$ is the ratio of rupture velocity to the limiting velocity $C_{lim}$ under a specific loading geometry (e.g., $\gamma=v_r/c_{Ral}$ for Mode II and $\gamma=v_r/c_S$ for Mode III, where $c_{Ral}$ and $c_S$ are the Rayleigh- and the shear-wave velocities, respectively).

\begin{figure}[h]
	\centering
	\includegraphics[width=0.7\linewidth]{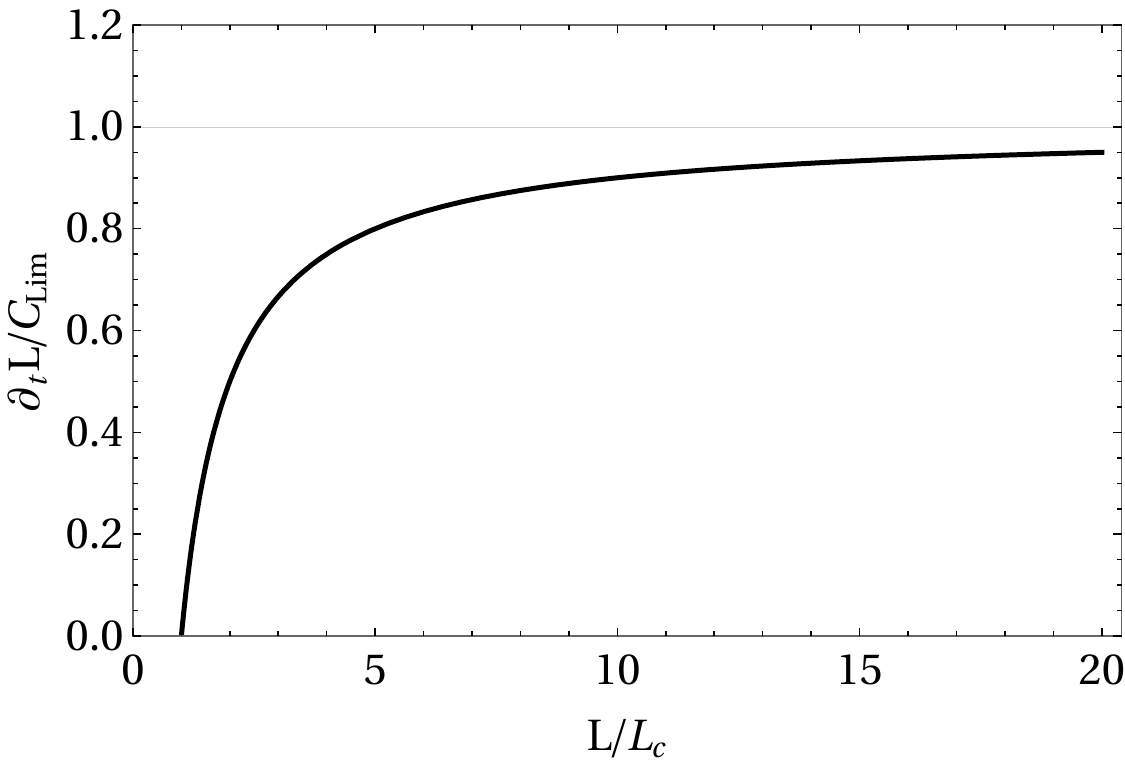}
	\caption{Rupture velocity as a function of position according to (\ref{eq:Ldiffeq}). Horizontal and vertical axis are normalised by characteristic length $L_c$ and velocity $C_{lim}$, respectively.  
	}
	\label{fig:v_L}
\end{figure}

\begin{figure}[h]
	\centering
	\includegraphics[width=0.7\linewidth]{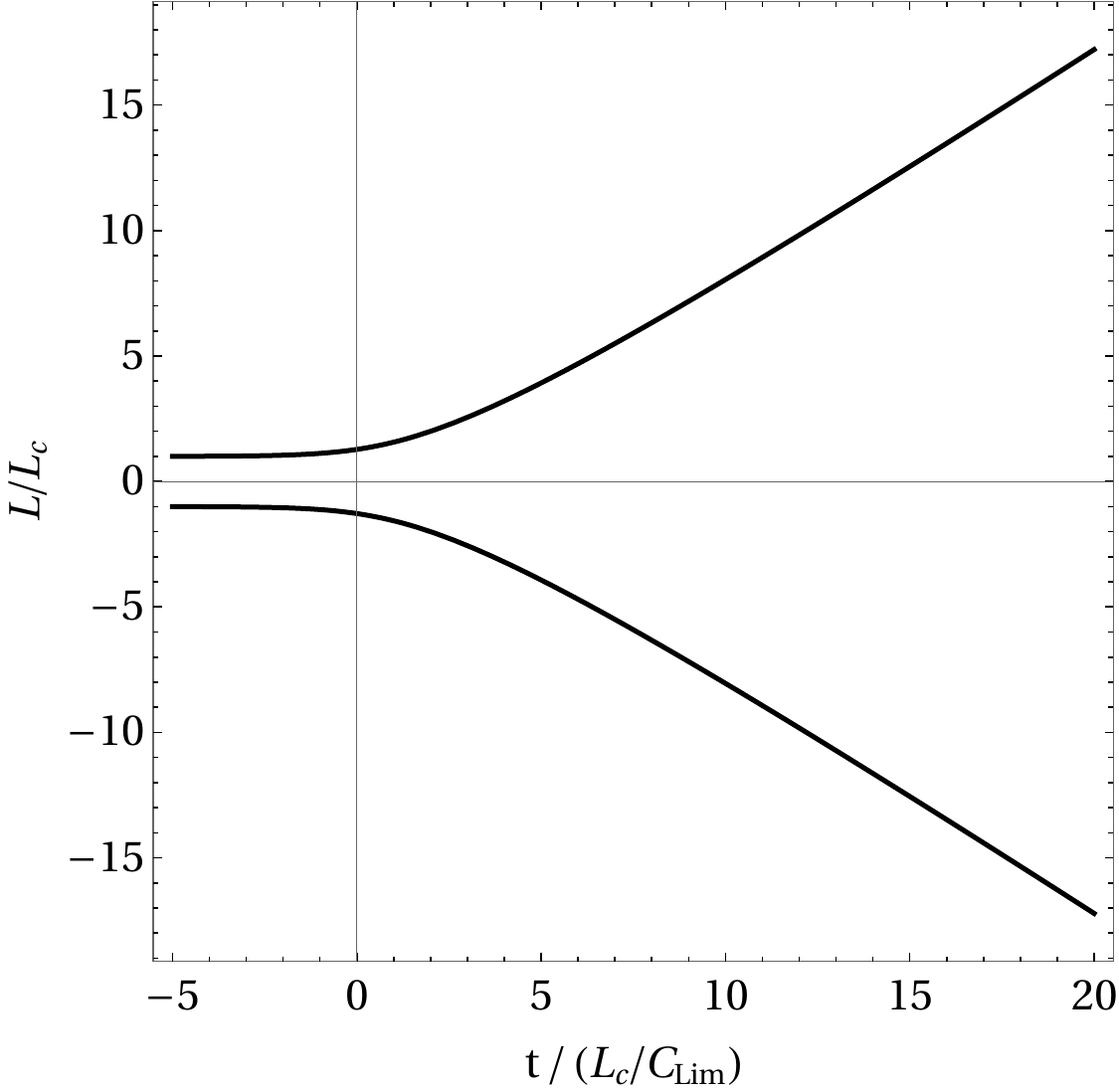}
	\caption{
		Rupture position as a function of time according to (\ref{eq:L_sol}) with $t_0 = 1$. Horizontal and vertical axis are normalised by characteristic length $L_c$ and time $L_c/C_{lim}$, respectively.  
	}
	\label{fig:L_t}
\end{figure}

A practical and general approximation for the crack tip motion equation (\ref{eq:eflow1}) can be rewritten as (see derivation in Appendix \ref{derivation}):
\begin{equation}
\Gamma \approx (1-\gamma)\ G_0
\label{eq:approx_flowdyn2}
\end{equation}
where $G_0$ is the static energy flow, and for a crack of length $L$:
\begin{equation}\label{eflow_static}
{\setstretch{2}
	\begin{array}{llc}
	G_0 &=& A_0\  \frac{\Delta\tau^2\ L}{\mu^\prime}
	\end{array}
}
\end{equation}
where $\Delta\tau$ is the stress drop inside the crack, $\mu'$ the shear stiffness and constant $A_0$ is a dimensionless shape factor involving loading geometry and Poisson ratio. 
In addition, the initial rupture patch dimension can be defined as the critical length of a Griffith crack \citep{Andrews1976}:
\begin{equation}
L_c =\frac{1}{A_0}\ \frac{\Gamma\, \mu'}{\Delta\tau^2}
\label{eq:Irwin}
\end{equation}
where $\Gamma = G_c + G_\textrm{off}$ is equivalent to the critical energy release rate $G_c$ in the traditional LEFM sense  \citep{Kammer2024}, plus any general non-local or off-fault energy dissipation $G_\textrm{off}$  \citep{Dunham2011,Xu2012a,Nielsen2016a,Aben2020}. If we derive $\Gamma$ from equation (\ref{eq:Irwin}) and divide it by $G_0$ as defined in (\ref{eflow_static}) we can write:
\begin{equation}
{\setstretch{2}
\frac{\Gamma}{G_0} =  \frac{L_c}{L}
\label{eq:ratios}
}
\end{equation}
and note that the ratio of energy dissipation $\Gamma$ to static energy flow will scale as the ratio of $L_c$ to $L$. Accessorily, at the quasi-static initiation of rupture propagation, (\ref{eq:ratios}) shows as expected that the energy flow $\Gamma$ is equal to the static energy flow for $L=L_c$. 
Substituting $\Gamma$ from the crack tip equation (\ref{eq:approx_flowdyn2}) into (\ref{eq:ratios}), 
and noting that $\gamma\ C_s = \partial_t L = V_r$, we can set the problem of rupture acceleration as the differential equation:
\begin{equation}
	{\setstretch{2}
	%yn'[t] == Clim/yc (1 - 1/yn[t])
	\frac{1}{C_{lim}}\,\frac{\partial L}{\partial t} = 
	\left( 1-\frac{L_c}{L}\right)
	\label{eq:Ldiffeq}
	}
\end{equation}
This extremely simple solution immediately illustrates (a) self-similarity in the ratio $L_c/L$, and that (b) rupture velocity at fixed $L$ is a decreasing function of $L_c$, in other words, larger break-out patches will accelerate more gradually.  
The rupture velocity solution $\partial_t{L}$ is shown in Figure (\ref{fig:v_L}) as a function of normalised time and length.

Furthermore, we can solve (\ref{eq:Ldiffeq}) for $L(t)$ to yield
\begin{equation}
% L(t) = L_c (1 + ProductLog[E^(-1 + Clim /yc (t + t0))]) 
L(t) = L_c 
\biggr(
1 + 
W_0
\left[
 	\exp
 	\left(
 	%\frac{C_{lim}}{L_c}(t - t_0) - 1  	
 	\frac{C_{lim}}{L_c}(t - t_0)
 	\right)
\right]
\biggr)
\label{eq:L_sol}
\end{equation}
as shown in Figure (\ref{fig:L_t}), where $W_0$ is the principal branch of the Lambert W function, and $t_0$ an arbitrary integration constant or reference time. 
The significance of $t_0$ is that $L=L_c$ is perfectly matched only at time $-\infty$, 
and that an hypothetical breakout patch infinitely close to instability would take an infinite time to accelerate in the absence of finite perturbations.

%Noting that
%\begin{equation}
%\lim\limits_{t\to\infty} W_0[\exp(a\ t)]=a\ t
%\end{equation} 
Note that the asymptotic behavior of (\ref{eq:L_sol})  at large time can be obtained as
\begin{equation}
\lim\limits_{t \to \infty} L(t) = L_c + C_{lim}\ (t-t_o)
\label{eq:limit}
\end{equation}
i.e., a constant rupture velocity at the limit speed $C_{lim}$, under the assumption that no supershear rupture transition has yet occurred.

Analytic solutions for unstable propagation in 3D are not known (e.g. a solution exists for the self-similar elliptical crack under constant rupture velocity, see \cite{Burridge1969}). However, approximation (\ref{eq:approx_flowdyn2}) can be used under the assumption that $C_{lim}=\sqrt{C_\textrm{Ral}\ C_s}$ is the geometric mean of the Mode II and Mode III subsonic limiting velocities, and that $L=\sqrt{A/\pi}$ where $A$ is rupture patch area. 
These assumptions are validated by a numerical simulation of 3D spontaneous rupture (see section \ref{numerical} below), where I also demonstrate numerically that the scaling in $L_c$ can be generalised to inhomogeneous parameter distributions on the fault and also applies non-trivial shapes of the initial breakout patch in 3D, where I define an equivalent radius, or average front position, as $L=\sqrt{A/\pi}$ where $A$ is rupture patch area. 

\begin{figure}[h]
	\centering
	\includegraphics[width=0.7\linewidth]{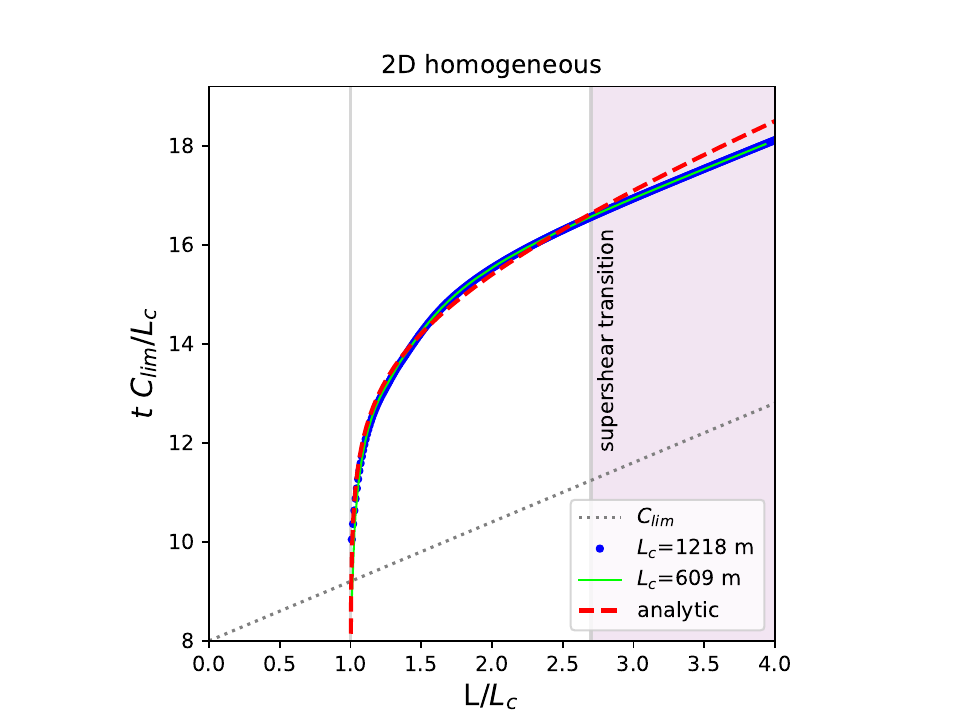}
	\caption{
		Comparison of the analytical self-similar solution (\ref{eq:Ldiffeq}) and the numerical simulation for a crack expanding under 2D in-plane loading geometry on a fault with homogeneous pre-stress. The
		scaling is illustrated by superimposing the simulation for two different initial rupture patches 
		($L_c=1218 \textrm{ vs. } 609$ m) after normalising the result by $L_c$. The analytical solution is very close to the numerical result, up until the transition to supershear velocity (supershear is not considered in the solution, as it is derived for the initial rupture acceleration in the sub-sonic break-out phase).   
	}
	\label{fig:2Dhomogeneous}
\end{figure}

\begin{figure}[h]
	\centering
	\includegraphics[width=0.7\linewidth]{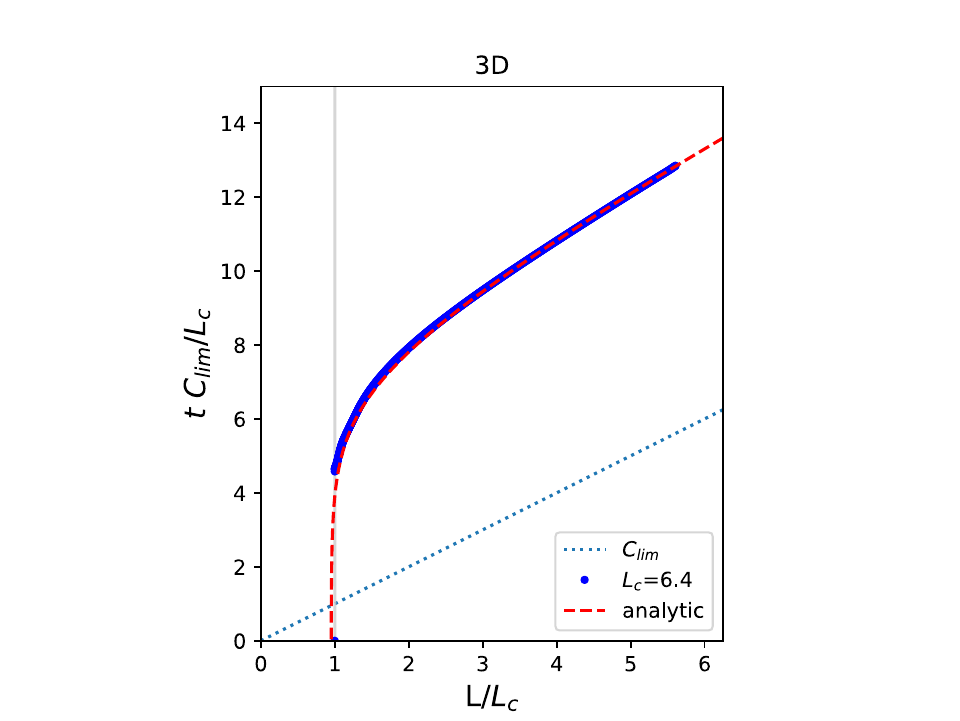}
	\caption{
		Comparison of the analytical self-similar solution (\ref{eq:Ldiffeq}) and the numerical simulation for a crack expanding in 3D on a fault with homogeneous pre-stress. Here, the equivalent length of crack is defined as $L=\sqrt{A/\pi}$ where $A$ is the area of the rupture in the numerical simulation. $L$ equates to radius for a circular rupture, or to the geometrical mean of long and short semi-axis in the case of an elliptical rupture.  Note that the analytical fit shown is obtained for $L=0.95$ in this case only: here the 0.95 fit is slightly better than for $L=1$ due to the numerical loading in the simulation, which slightly overshoots the critical load for $L=1$). 
	}
	\label{fig:3Dhomogeneous}
\end{figure}

\begin{figure}[h]
	\centering
	\includegraphics[width=0.49\linewidth]{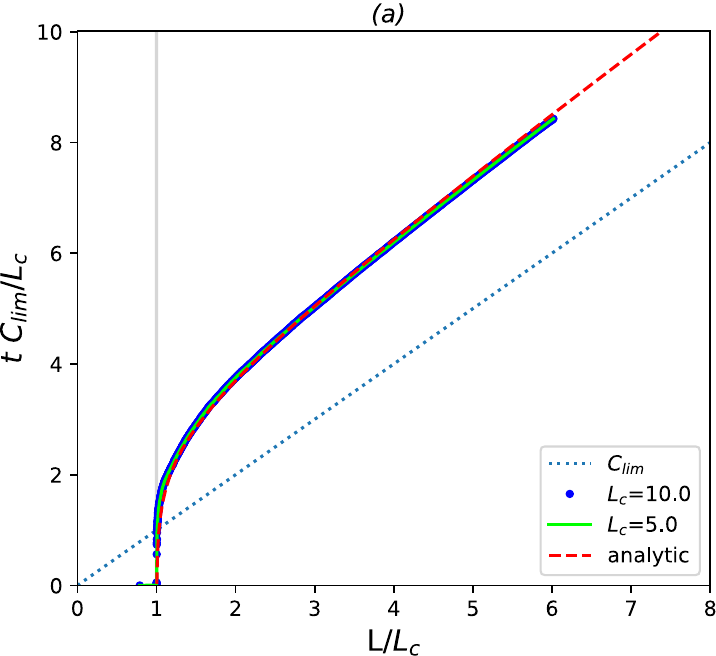}
	\includegraphics[width=0.49\linewidth]{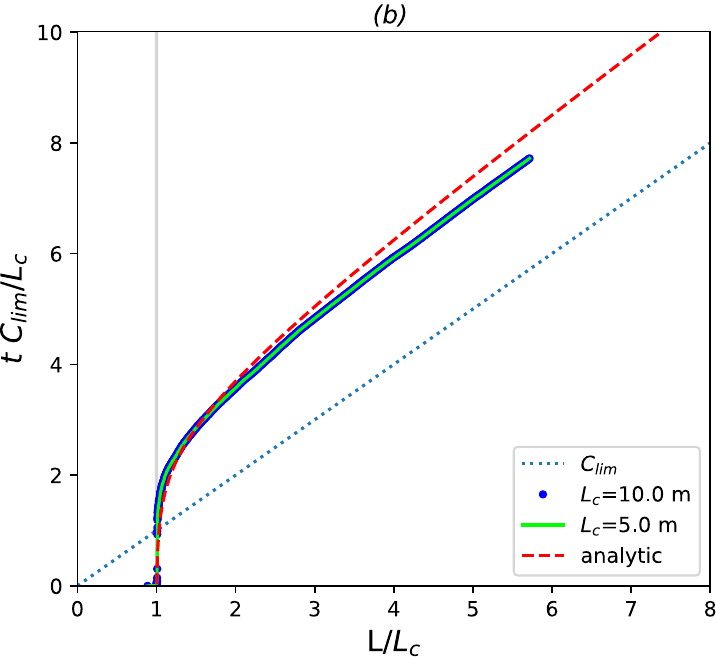}
	\caption{
		Comparison of the analytical self-similar solution (\ref{eq:Ldiffeq}) and the numerical simulation for a crack expanding in 3D on a fault with (a) mildly inhomogeneous pre-stress and (b) strongly inhomogeneous prestress with wavenumber powerlaw in $k^{-1.2}$. To illustrate the generality of the scaling relation obtained in (\ref{eq:Ldiffeq}), two different initial lengths $L_c=5, 10$ m were simulated, under inhomogeneous pre-stress, and using a non-elliptical rupture shape for the initial rupture patch. (For the general rupture shape, rupture length $L$ is defined as in Fig. \ref{fig:3Dhomogeneous}). In case (b) the super-shear transition takes place earlier in the numerical simulation, due to patches of increased prestress, so that the rupture front (t, L) curve deviates from the analytical solution at 2-3 s.
	}
	\label{fig:3Dinhomogeneous}
\end{figure}

\section*{Identifying the self-similar $L(t)$ solution in numerical simulations and experiments \label{numerical}}

\subsection*{2D in-plane solution}

I numerically solved the spontaneous rupture propgation in 2D, for in-plane stress loading, using a finite-differences, stress-velocity formulation on a staggered grid  \citep{Virieux1986,Festa2003} of $nx \times ny=1024 \times 516$ nodes. A fault line 
was defined in the model along the $x$ direction, with the conventional slip-weakening friction formulation  \citep{Ida1972b,Andrews1976,Tinti:2004} such that
\begin{equation}
{\setstretch{2.0}
\tau =	\left\{
\begin{array}{cc}
(\tau_p - \tau_r)\frac{U}{D_c} + \tau_r   & \textrm{ for } U < D_c \\
\tau_r                                    & \textrm{ for } U \ge D_c
\end{array}
\right.
}
\label{eq:slip_weak_fric}
\end{equation} 
applied at any point on the fault after the peak $\tau = \tau_p$ stress has been reached.

Rupture was initiated from a patch of length (1) $L_c = 609.375\textrm{ m}$ (195 grid nodes with spatial step dx=3.125) and (2) $L_c = 1218.75\textrm{ m}$ (195 grid nodes with dx=6.25 m). In both cases, shear stress $\tau_{xy}$ was slowly increased until rupture started to propagate spontaneously.

The resulting rupture front position (Fig. \ref{fig:2Dhomogeneous}) is shown as a function of time in Fig. \ref{fig:2Dhomogeneous} for both values of $L_c$, along the analytical solution obtained in equation (\ref{eq:L_sol}) and $t_0=14.4 \textrm{s}$, with suitable normalisation of time and position. 

The numerical solutions for the different $L_c$ values are matching perfectly, 
illustrating the self-similarity predicted by the analysis. 
Within the breakout phase and throughout the predicted regime of application (subshear rupture velocity), 
the analytical and numerical results also indicate a very good match, and start to diverge mainly when the supershear transition is initiated (at about $L/L_c = 2.7$).

\subsection*{3D solutions with {area} -- $L^2$ equivalence}

Although solution (\ref{eq:L_sol}) is obtained for 2D translational symmetry, the energy flow essentially depends on the characteristic dimension $L$ of the crack under any geometry. 
What is then a sensible estimate of such length $L$ for a general crack shape? 
Here I define $L$ by assuming that the area of $A$ of an expanding rupture increases as $L^2$ and use the approximation of circular rupture where 
\begin{equation}
L=\sqrt{A/\pi}.
\label{eq:length-area}
\end{equation}
For example, assuming that the non-equant shape of an expanding crack is an ellipse (a fair approximation for a sub-sonic crack in a homogeneous medium, see  \cite{Burridge1969, Favreau2002}) the area is $A=\pi\ a_e\ b_e$ and in this case $L=\sqrt{A/\pi}=\sqrt{a_e\ b_e}$ is the geometrical mean of the ellipse major and minor axes $(a_e, b_e)$, respectively.
The limit subsonic velocity $C_{lim}$ is Rayleigh wave velocity and shear wave velocity in the in-plane and anti-plane crack front directions, respectively, so the expected average
limit velocity is the geometric mean $C_{lim}=\sqrt{C_\textrm{Ral}\ C_s}$. The length at any time (including $L_c$) can be obtained from (\ref{eq:length-area}).

The general 3D equivalent $C_{lim}$ and $L$ values defined above can be substituted in solution (\ref{eq:L_sol}) for comparison with numerical results obtained for any crack geometry, including a crack that is initiated with a non-elliptical geometry and propagatges in a non-homogeneous medium, to check to what extent solution (\ref{eq:L_sol}) can be generalised to arbitrary crack shapes.  

Such comparison is shown  in figures  (\ref{fig:3Dhomogeneous},\ref{fig:3Dinhomogeneous})
for an initial crack with a non-trivial geometry (non-elliptical shape of a quatrefoil) and for propagation both under homogeneous and inhomogeneous pre-stress.

\subsection*{Dynamic rupture in laboratory experiments}

Observation of the dynamic rupture front can be achieved in the laboratory using high-speed photographic interferometry. Rupture can either be initiated by an artificial impulsive trigger \citep{Xia2004}, or left to grow as a spontaneous instability under loading \citep{Nielsen2010a}; in the latter case, a circular buffer acquisition system allows to monitor the nucleation and break-out phase preceding the fully dynamic, close-to-sonic rupture propagation \citep{GuerinMarthe2019}. 

\begin{figure}[h]
\centering
\includegraphics[width=0.7\linewidth]{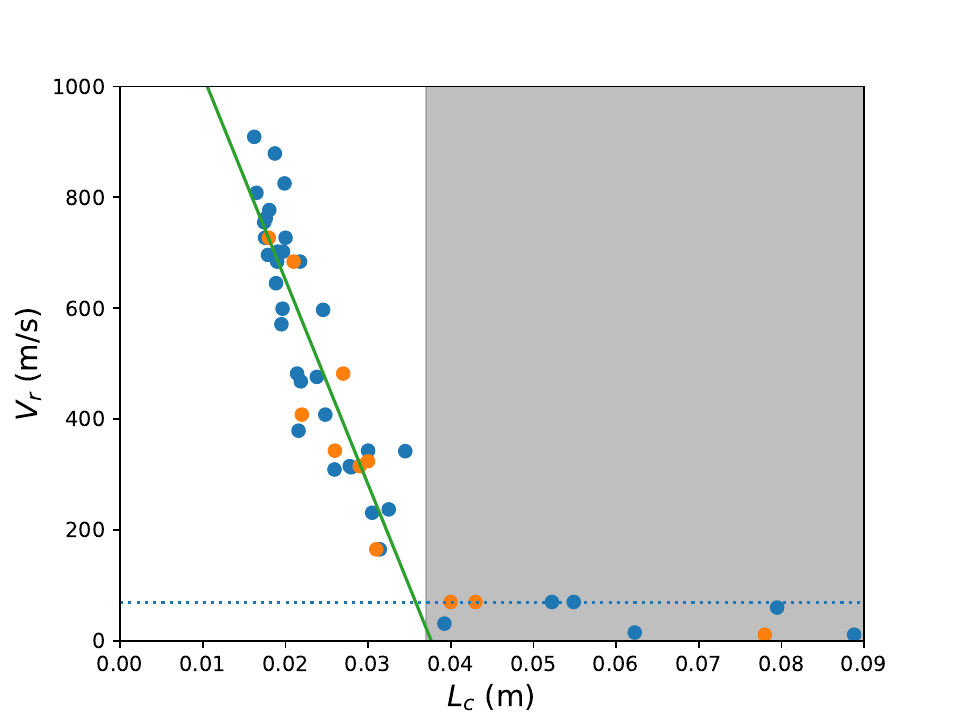}
\caption{
	Comparison of analytical self-similar solution of equation (\ref{eq:Ldiffeq}) and laboratory experimental results from  \citep{Latour2013}. 
	A linear decreasing trend is expected for $V_r$ versus $L_c$, where $V_r$ is measured at a fixed distance $L$ from the nucleation patch. 
	Orange dots represent experiments
	where $L_c$ was directly measured on the interferograms; blue dots represent experiments where $L_c$ was inferred. The green line
	shows the fit with $C_{lim}=1390$ m/s (standard deviation of 95 m/s), $L=0.037$ m . The greyed-out area shows results where the velocity is 
	sufficiently low ($V_r\leq 5\% \ C_{lim}$) to consider a quasi-static  
	process where the dynamic solution (\ref{eq:Ldiffeq}) does not apply. See text for further details.
}
\label{fig:Latour_Vr_Lc}
\end{figure}

Nucleation and gradual acceleration of the rupture front to sub-sonic dynamic velocity were monitored in a series of experiments  \citep{Latour2013}, providing useful data to verify the $L$ vs. $L_c$ similarity. 
The rupture front position was retrieved from the high-speed photography images. It was also possible to measure $\partial_t L(L)$ using piezo-electric sensors along the fault line. The nucleation patch was in approximately the same position along the fault for all ruptures regardless of variation of $L_c$, so that $L$ can be considered as constant in first approximation. We use those results here to illustrate the observed $\partial_t L, L_c$.  According to equation \ref{eq:Ldiffeq}, $v_r=\partial_t L$ is expected to be decrease linearly with $L_c$, which is compatible with the experimental results shown in Figure (\ref{fig:Latour_Vr_Lc}).

\section*{The case for a correlation between $L_f$, $L_c$ and final magnitude \label{case}}

I develop below a simplified model based on rupture energy concepts; the aim is not to integrate the complete and advanced rupture features, but rather to illustrate the possible scaling of rupture initiation on an inhomogeneous fault through a bare-bones dimensional analysis.
 
Assume a locked fault under inhomogeneous initial stress ${\tau}_{1}(\textbf{x})$, with the potential to yield and undergo frictional weakening to $\tau_r(\textbf{x})$.
Consider failure of a patch $P$ of the fault with area $L\times L$, 
undergoing the average stress drop
$\Delta\tau(P)=(1/L^2)\ \int_{L^2} \tau_1(\textbf{x}) - \tau_r(\textbf{x}) \ dx$ and an 
average dissipation $\Gamma(P)$. After re-arranging equation (\ref{eq:Irwin}) and taking the square root, we can write the criterion that rupture may propagate 
from patch $P$ provided that the dimensionless quantity $\zeta$ exceeds 1, or:
\begin{myequation}{\label{eq:minstress}}
\zeta(P) &\geq  1 & \\
 & \textrm{with} & \\
\zeta(P) & = &\chi(P)\    {L(P)}\\
\chi(P) & = & {\frac{A_0}{\mu'\Gamma(P)} }\    \Delta\tau^2(P)
\end{myequation}
\begin{figure}[h]
	\centering
	\includegraphics[width=0.7\linewidth]{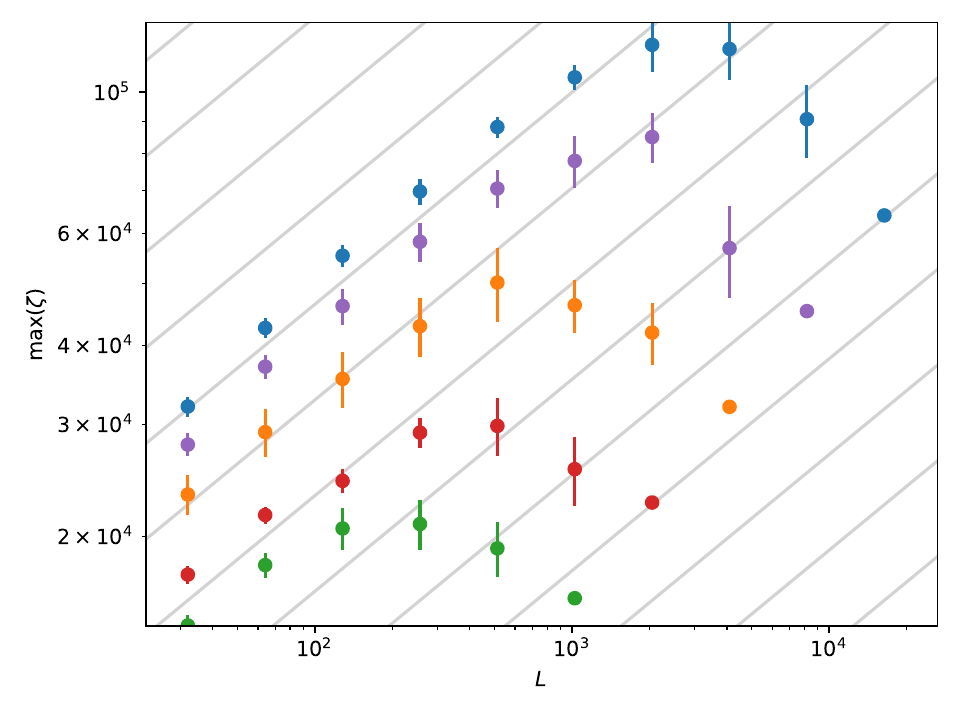}
	\caption{
		Maximum value of $\zeta$ (where $\zeta\propto{\Delta\tau}^2{L}$ see equation \ref{eq:minstress}) on patches of different size $L$ on a fault $L_f \geq L$. Three different upper fault dimensions $L_f=1024$, 2048, 4096, 8192 and 16384 are shown (green, red, orange, purple and blue, respectively). An inhomogeneous, self-affine distribution of $\chi$ is assumed on the fault. The rupture will break-out from the patch with maximum value of $\zeta = \chi \ L$ according to the rupture criterion of equation (\ref{eq:Irwin}). The size of the break-out patch (maximum  $\zeta$) increases with $L_f$, the maximum fault dimension. See text for further details. The drop at high $L$ values is due to $L$ approaching $L_f$ and the consequently the maximum value approaching the average value of the whole fault.  See text for further details. 
	}
	\label{fig:tau_sqrt_L}
\end{figure}

\begin{figure}[h]
	\centering
	\includegraphics[width=0.7\linewidth]{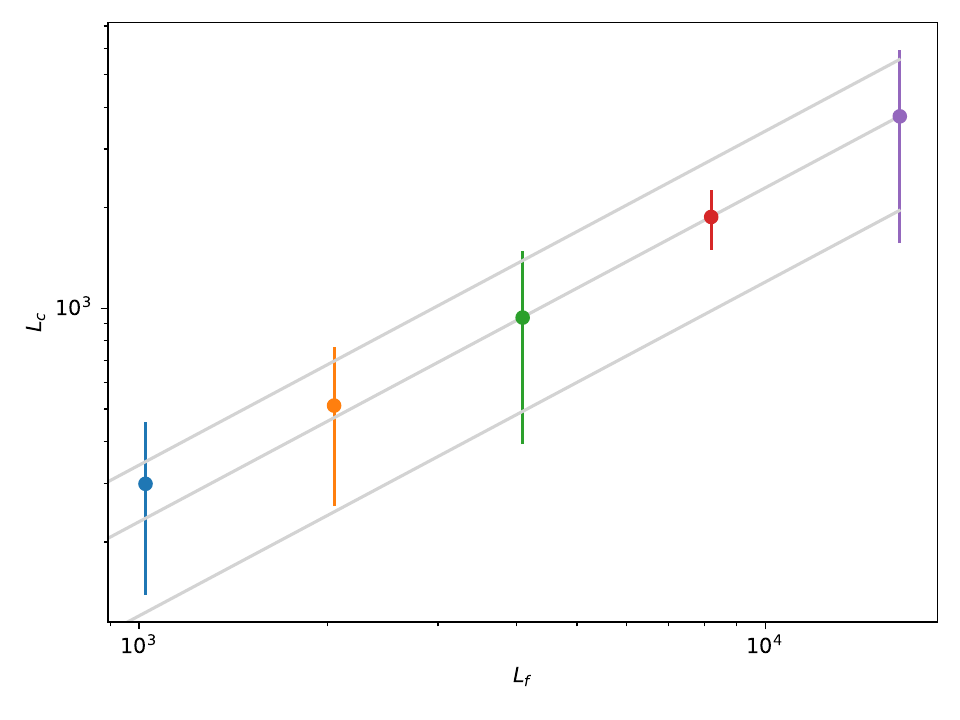}
	\caption{
		Mean value (dots) of $L_c$ and standard deviation (bars) obtained for all simulations, as a function of the maximum fault size $L_f$. (The $L_c$ value in each simulation corresponds to the size of sub-fault with maximum $\zeta$ according to equation (\ref{eq:minstress}). 
		The three grey lines indicate the best linear fit, and plus and minus the error in slope, according to $L_c/L_f = 0.23 \pm 0.11$.
	}
	\label{fig:Lf_Lc_linearfit}
\end{figure}

Under increasing tectonic load, rupture breakout will start most likely from the fault patch with the largest possible $\zeta$ value. In the trivial case of homogeneous parameters of $\chi$, $\zeta$ is maximised for the largest $L$, therefore the fault dimension $L=L_f$. However, natural faults are highly inhomogeneous; therefore, it is understood that a patch $L_c \ll L_f$ can indeed induce breakout if it is much weaker than the surrounding fault segment $L_f$.
Hence the expected dimension of $L_c$ relative to $L_f$ will depend on the spatial probability density of fault weakness (in the broad sense) and its spatial distribution. The expected $L_c$ is determined by the probability $P(L_1, L_2)$ of drawing a sample $L_c$ that verifies (\ref{eq:minstress}) within an interval $[L_1, L_2]$ based on the distribution of combined fault parameters.  However, for a given fault segment
we will have $P(L>L_f)=0$. The upper bound of a distribution will affect the mean value, therefore statistically speaking larger $L_c$ values will be found on larger fault segments. Simply put, large nucleations cannot occur on small faults. Therefore large nucleations are indicative of large faults where potentially large earthquakes occur.  
To illustrate this point, I build the following elementary model. 

One main sources of inhomogeneity is the fault surface roughness which is well described by a self-affine distribution with a Hurst exponent $H\approx 0.6$ parallel to slip and $H\approx 0.8$ perpendicular to slip direction \citep{Candela2012}. 
The roughness distribution governs the normal stress distribution through the elastic contact across the fault surface, resulting in a Hurst exponent $H'=H-1$ \citep{Hansen2000}. Arguably, the same exponent $H'$ will control the stress drop $\Delta\tau$ and the dominant features of the $\chi$ distribution. 

I investigate here a possible relation between 
$L_c$ and $L_f$ assuming that the combined parameters in $\chi$ follow a self-affine of
Hurst exponent $H'=-0.2$, corresponding to a power spectrum distribution with slope $\beta=-2-2 H'$ \citep{Hansen2000}. To this end I populated a square $N\times N$ array with Gaussian white noise with standard deviation one fifth of the mean (e.g. mean 500 and std 50). I then took the Fourier transform of the noise array and multiplied the Fourier coefficients $C_{pq}$ by a powerlaw of the wavenumber module $k=\sqrt{p^2+q^2}$ with exponent $\beta/2 = -1 - H'=-1.2$, so that $C_{pq}^{filt}=C_{pq}\ k^{-1.2}$, following a similar procedure as described in \citep{Turcotte1997} with minor variations. 
Finally, I took the inverse Fourier transform to obtain a random $\chi$ distribution with the desired self-affine properties.

Assuming that the maximum size of a fault is $L_f=N$, I used five different array sizes of dimensions $N = 1024$, 2048, 4096, 8192, 16384, and repeated the procedure six times with different random distributions, thus creating 30 arrays in total.
I then split each of the arrays of size $N\times N$ in sub-arrays $n\times n$ where $n=\nicefrac{N}{2}, \nicefrac{N}{4}, \nicefrac{N}{8}$ \dots down to a minimum size $n=32$, assuming $L=n$ is the size of a sub-fault within $L_f$ (see illustration in Supplementary Information).  

Within each of the sub-arrays, considered as a potential fault failure patch of size $L$, I computed the $\zeta$ value, as shown in Figure (\ref{fig:tau_sqrt_L}). The mean and standard deviation of $\zeta$ are shown for the six realisations, for all five $L_f$ and all sub-arrays sizes $L$.
The value of $L$ where $\zeta$ peaks for a given $L_f$ indicates the most likely dimension of failure patch $L_c$, and such value appears to scale with $L_f$. 

I then computed the best linear fit between $L_c$ and $L_f$ for all models, 
resulted in  $L_c / L_f = 0.23 \pm 0.11$ as shown in Figure (\ref{fig:Lf_Lc_linearfit}). The most frequent breakout patch $L_c$ is about one fifth of the maximum fault size, for the specific model implemented here.

\section*{Moment rate indicative of $L_c$}

\begin{figure}[h]
	\centering
	\includegraphics[width=0.7\linewidth]{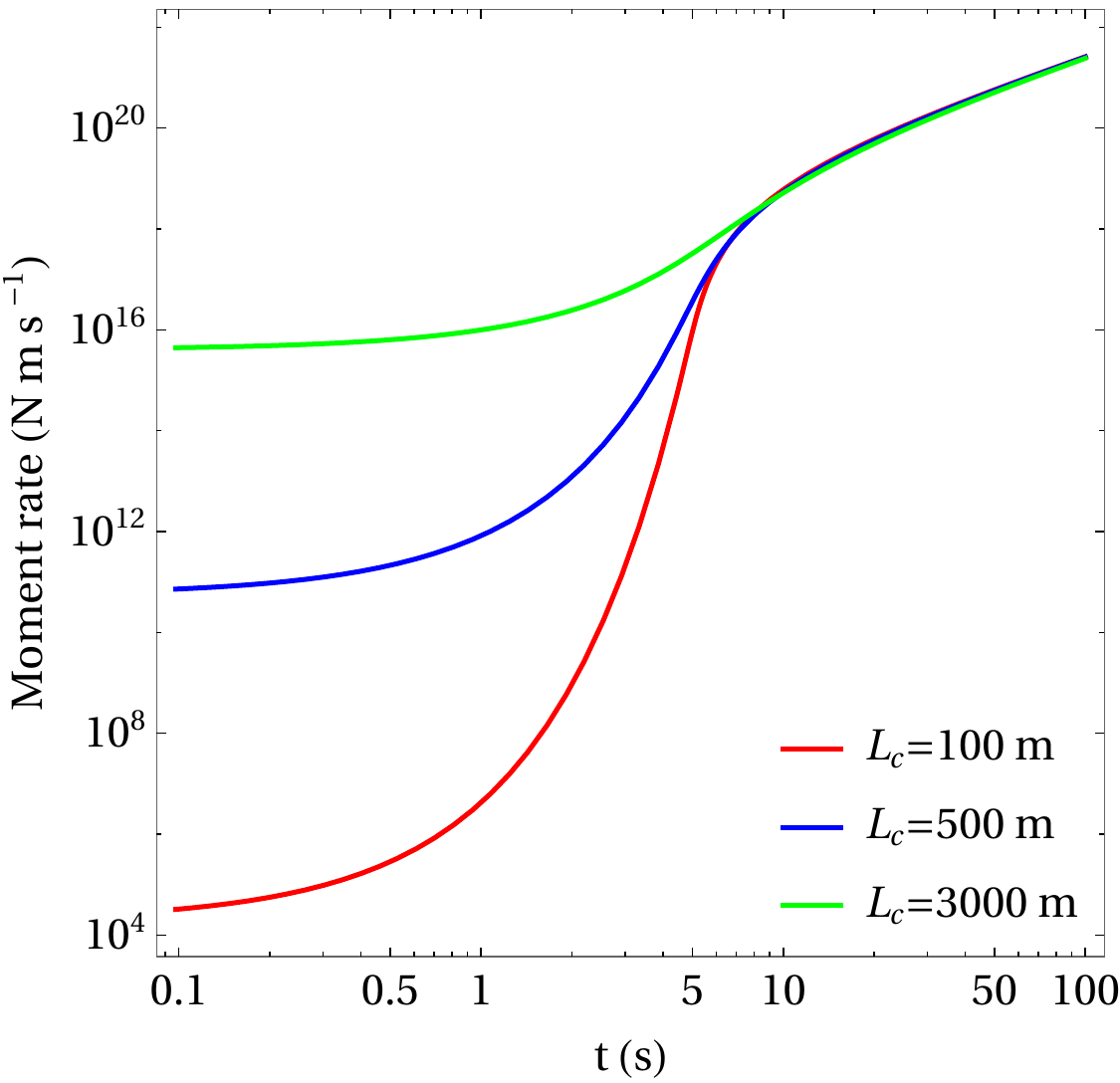}
	\caption{
		Moment rate according to equation (\ref{eq:momentrate}) for three different initial rupture sizes $L_c=100,\ 500$ and 3000 m, for indicative parameters $\Delta\tau=5\textrm{ MPa}$ and $C_{lim}=2700\textrm{ m/s}$. The moment rate differs remarkably in the first few seconds of rupture (about 5 s in the above examples, and up to about 10 s for $L_c$ of more than 10 km). The moment rates then converge as the rupture expansion velocity approaches the asymptotic value $C_{lim}$. 
	}
	\label{fig:momentrate}
\end{figure}

For a rupture expanding as equation (\ref{eq:L_sol}), the moment rate can be obtained (see Supplementary Information) resulting in:
\begin{equation}
	\dot{M}_o (t)= 3C\  C_{lim}\ \Delta\tau\ L_c^2 \left( W_o[e^{\frac{C_{lim}}{L_c}(t - t_0)}] ^2 + W_o[e^{\frac{C_{lim}}{L_c}(t - t_0)}] \right).
	\label{eq:momentrate}
\end{equation}
which is illustrated in Fig. (\ref{fig:momentrate}) for three different initial break-out sizes $L_c$, using indicative parameters 
$\Delta\tau=5\textrm{ MPa}$ 
and 
$C_{lim}=2700\textrm{ m/s}$. 
The  initial seconds of rupture indicate a big difference in the moment rate for different initial lengths, but after about 5 s all solutions converge and to an identical asymptotic moment rate (see asymptotic solution in \ref{eq:limit}) where the square term dominates  such that

\begin{equation}
\begin{array}{ll}
\lim\limits_{t \to \infty}
\dot{M}_o (t) & = 3C\  C_{lim}\ \Delta\tau\ L_c^2  \left( \frac{C_{lim}^2}{L_c^2}\ (t - t_0)^2 + \frac{C_{lim}}{L_c}\ (t - t_0) \right) \\
                     & \approx 3C\  C_{lim}\ \Delta\tau\  C_{lim}^2\ (t - t_0)^2  
\end{array}
\end{equation}
Hence the initial seconds of the moment rate can be indicative of the initial rupture size. 

The far-field displacement is proportional to the moment rate such that
\begin{equation}
u(t) = \dot{M}_o(t-\tau,\theta,\phi) \ \frac{e^{-\pi \omega r / c Q}\ \mathfrak{R}_{\theta\phi}}{4\pi\rho c^3 r},
\label{eq:farfield}
\end{equation}
where $r$ is the source-station distance along the ray path, $c$ the wave velocity (either P- or S-wave depending on the considered phase), 
$\rho$ is mass density. The angles $\theta,\phi$ describe the receiver position with respect to the fault plane and slip direction. Here $\dot{M}_o(t-\tau,\theta,\phi)$ is the apparent moment rate where directivity effects are accounted for 
\citep{Venkataraman2004}, 
which reduces to $\dot{M}_o(t-\tau)$ for a point-source approximation \citep{Aki2002}. $\mathfrak{R_{\theta\phi}}$ is a coefficient accounting for the radiation pattern. The exponential term accounts for attenuation ($\omega$ is frequency, $Q$ is the quality factor). 

Relation (\ref{eq:farfield}) is used either implicitly or explicitly in a number of 
studies to infer moment rate (also called source-time function) from far-field seismic displacement recordings %\citep{Ishihara1992,Vallee2010,Ye2016,Colombelli2014,Meier2017,Colombelli2020}. 
Assuming that moment rate function is inferred from initial seconds of far-field displacement seismogram according to (\ref{eq:farfield}), moment rate can then be used to infer the initial dimension $L_c$ of the break-out rupture according to (\ref{eq:momentrate}). Note 
that in the far field radiated wavefield from an extended source, the initial seconds of a seismogram  may be indicative of later rupture times \citep{Murphy2009}, as
illustrated by the theory of isochrones 
\citep{Spudich2004} 
in the high frequency limit; in addition as noted above the amplitude of motion is affected by directivity and receiver position relative to the focal mechanism. Therefore suitable corrections or averaging over many stations is required for an accurate estimate.

In terms of rupture front acceleration, interesting indicative numbers 
can be derived by comparing the behaviour of initial ruptures of, for example, 300 m, 3 km and 30 km.  First from equation (\ref{eq:Ldiffeq}) we can derive the rupture tip position $L_\phi$
at which a fraction $\phi$ of the limiting rupture velocity has been reached:
\begin{equation}
L_\phi= \frac{L_c}{1-\phi}
\end{equation}  
then replacing $L_\phi$ in equation (\ref{eq:L_sol}) we obtain the corresponding time $t_\phi$ 
\begin{equation}
t_\phi = t_0 + \frac{L_c}{C_{lim}} \log \left(\frac{\phi\ e^{\frac{\phi
			}{1-\phi }} }{1-\phi
	}\right)
\end{equation}
Taking the difference between the time where rupture tip is propagating 
at $\phi=0.1\ C_{lim}$ (as indicative of velocity where wave radiation 
is starting to be noticed) and the time where it has reached $\phi = 0.86\ C_{lim}$
(a value close to the maximum sub-sonic rupture velocity), we obtain $\Delta t \approx 10\ L_c/C_{lim}$. Assuming $C_{lim}=3$ km/s it will take  approximately 0.33, 1, 10, and 100 s to transition from $V_r=0.3$ km/s to 2.58 km/s  for an initial breakout length of 0.1, 0.3, 3 and 30 km, respectively.  
Thus there is a substantial difference (from 0.33 s to 100 s) in the duration of acceleration  between a modest (100 m) and a large breakout dimension (30 km), which should, in principle, 
be readily detected from seismological data.

\section*{Instability and rupture propagation under Rate and State friction}

The rupture breakout phase can be described as the start of the dynamic rupture propagation, 
where the inertial terms and wave radiation are no more negligible and high velocity frictional weakening starts. 

The breakout follows a possibly long preparatory stage where quasi-static instability develops at low slip velocity and leads to rupture nucleation. Such stage is best described in the framework of rate-and-state friction and has been the object of numerous modelling studies. 

It is generally considered that instability can occur under rate-and-state friction in situations where $a-b < 0$ and the equivalent  stiffness $K_f$ of the sliding system (e.g., bloc slider or fault) is smaller than the critical stiffness $K_c = (b-a)\sigma / D_c$ (for negligible inertia), as argued in  %\cite{Rice1983a}. 
$a$ and $b$ are dimensionless parameters controlling direct and delayed response to a step in slip velocity, $D_c$ is the characteristic slip distance over which the delayed response occurs, and $\sigma$ is the normal stress. Fault stiffness can be defined as the ratio of shear modulus to fault characteristic length $L$ such that $K_f = C \mu'/L$ ($C$ is a geometric factor depending on loading geometry in 2D and rupture shape in 3D, e.g. $2\times 7\pi/16$ for a circular fault). Equating fault stiffness to critical stiffness, and assuming $C=1$, results in a minimum length of unstable fault patch %\citep{Rice1983a} 
\begin{equation}
L_{RR} = \frac{\mu' D_c}{(b-a)\sigma}.
\end{equation}
that accurately describes instability close to the steady-state (i.e. regime where $V \theta/ D_c \approx 1$) where $\theta$ is a state variable with time dimensions. However,
as noted by \citep{Dieterich1992} already, for regimes well above steady state ($V \theta/ D_c \gg 1$) the instability length under an ageing law is more accurately matched by the smaller value of
\begin{equation}
L_{b} = \frac{\mu' D_c}{b\sigma}.
\label{eq:L_b}
\end{equation}

$L_{RR}$ can be generalised \citep{Rubin2005,Ampuero2008} to consider sections of the fault that are not close to the steady-state, as expected at the tips of a propagating rupture where an abrupt change in stress / slip velocity would locally set the fault far above  the steady-state (i.e. a regime where $V \theta / D_c \gg 1$). The conclusion is that some variability in the instability length can be expected within identified end-members, as shown in either numerical and laboratory experimental ruptures \citep{GuerinMarthe2019}.

Propagation of rupture under rate-and state can be re-framed as an energy criterion, by 
using equation (\ref{eq:Irwin}) and replacing the expected stress changes under rate-and-state
friction. Noting that the velocity jump at the propagating rupture tip is sufficiently abrupt, it can be assimilated to a velocity step typically used in friction tests, with a response proportional to the step in log of velocity, i.e. a quasi-instantaneous stress rise $a\ \sigma \Delta(\ln V)$ followed by a breakdown drop $\tau_p - \tau_r = b\ \sigma\ \Delta(\ln V)$ from peak value $\tau_p$ to relaxed value $\tau_r$, over a characteristic slip distance $\delta_c$, and a final stress drop from initial to relaxed value $\Delta\tau = (b-a)\sigma\ \Delta(\ln V)$. Following \citep{Rubin2005} we note that the equivalent weakening distance for large velocity step (under the aging law) is $\delta_c = D_c (\tau_p-\tau_r)/b$. Substituting the above approximate values of $\tau_p - \tau_r$, $\Delta \tau$ and $\delta_c$ into equation (\ref{eq:Irwin}), and assuming that the frictional dissipation can be approximated as 
$\Gamma_{friction} = \frac{\delta_c}{2} (\tau_p - \tau_r) $, we obtain:
\begin{myequation}{\label{eq:Linf}}
	L_c &\approx &\frac{1}{2 A_0}\ \frac{D_c\ \mu'}{\sigma} \frac{b}{(b-a)^2} \\
	&    =   & L_{\infty}
\end{myequation}
Note that the above value $L_c = L_\infty$ was initially proposed in \citep{Rubin2005}, 
under a series of approximations. Such approximations were justified in view of the results of numerical simulations conducted under the aging law and 2D loading geometry resulting in $A_0=\pi/2$. 

In light of equation (\ref{eq:Linf}) we may seek an alternative value of $\zeta$ under
rate and state friction, where 

\begin{equation}\label{eq:minstressras}
\begin{array}{ccccc}
\zeta_{RaS}(P) &\geq  1 & & &\\
%\zeta_{RaS}(P) & = &\sqrt{\frac{2 A_0}{\mu'\sigma D_c b} }\    \sigma (b-a)\sqrt{L}\\
%\zeta_{RaS}(P) & = &\sqrt{\frac{2 A_0}{\mu'\Gamma} }\    \Delta\tau \sqrt{L}
\zeta_{RaS}(P) & = &{\frac{2 A_0}{\mu'\sigma D_c b} }\    \sigma^2 (b-a)^2 {L}
& =&{\frac{2 A_0}{\mu'\Gamma} }\    \Delta\tau^2 {L}
\end{array}
\end{equation}
With these remarks it becomes clear that the scaling of $\zeta_{RaS}$ and $\zeta$ are
equivalent in the sense that
\begin{myequation}{\label{eq:equiv}}
	\frac{\Delta\tau^2}{{\Gamma}}=\frac{\sigma^2 (b-a)^2} {{\sigma D_c b}}
\end{myequation}
If the inhomogeneity of the combined parameters ${\sigma^2 (b-a)^2}/{{\sigma D_c b}}$ on the fault surface follows a similar distribution (e.g. self-affine) as that controlling $\Delta\tau^2/{\Gamma}$ in the dynamic regime, then similar remarks on the rupture breakout scaling apply under the rate and state regime as under the fully dynamic regime as described above 
relative to the distribution of $\zeta$. It has been argued that the large nucleations reported in some subduction-zone earthquakes are due to small values of $(b-a)$ and linked to the presence of fluids in the fault zone \citep{Cubas2015}. This argument is also compatible with the definition of $\zeta_{RaS}$ where $(a-b)$ appears, and still requires that the nucleation is supported by a large fault segment.

As rupture velocity increases, rupture transitions from the quasi-static nucleation process (with negligible inertia and no seismic wave radiation) to the fully dynamic seismic stage. When slip velocity reaches a few cm/s, enhanced velocity weakening processes are expected to kick-in \citep{ditoro2011,Spagnuolo2016,Nielsen2021}, where both friction and fracture energy will obey a different regime.  An empirical law that encapsulates both rate and state and high velocity weakening was proposed for rupture numerical models of \citep{Zheng1998}; a similar form was proposed to fit experimental data \citep{Spagnuolo2016}, who discussed the variation in critical stiffness due to the triggering of enhanced weakening. 

Due to the enhanced weakening, both fracture energy and stress drop are expected to be larger in the dynamic regime. The ratio ${\Gamma}/{\Delta\tau^2}$ before and after the transition will determine whether the effective $L_c$ is larger or smaller in the dynamic (enhanced weakening) stage or in the quasi-static (rate and state weakening) stage. 

The transition between the quasi-static and dynamic regimes may result in a variety of
nucleation-breakout sequences. 
If the dynamic $L_c$ is larger than that the static one, the rupture may continue to propagate under rate and state friction, likely at a modest rupture velocity and with minor stress drop, until $L_c^{dyn}$ is reached. In the laboratory, this instance may correspond to
the observation of a slow stable front at $\approx 5 \% C_s$ \citep{Nielsen2010a} or
the detachment fronts observed by \citep{Rubinstein2004} before the dynamic rupture fully develops. The observed stable front can be reproduced numerically under rate and state law under a favourable set of parameters  \citep{Kaneko2011}. 
In nature, observations compatible with this behaviour have been reported, for example, slow slip transients propagating toward the initial rupture point \citep{Kato2012} for the Tohoku, 2011 subduction thrust earthquake off the East coast of Japan. 
If, on the contrary, the dynamic $L_c$ is smaller than the quasi-static one, 
it is expected that a large region will slip initially, followed by the coalescence of 
a smaller dynamically unstable patch within it, as suggested by reports of the
extended slow slip event followed by the Valparaiso (2017) earthquake \citep{Ruiz2017,Moutote2023}.

\section*{Conclusions}

I have identified a self-similar scaling in the acceleration of rupture propagating from an initial crack based on fracture energy arguments, showing that the rupture front position 
$L$ scales as the critical crack length $L_c$ in space and as $L_c/C_{lim}$ in time.  
I derive an analytical solution for the rupture front position using an approximate velocity dependence of energy flow.  
Using finite difference numerical solutions, I show that the self-similar scaling remains valid for the fully elastodynamic problem where no approximation is made, both in 2D and 3D, and that the approximate solution is very close to the full solution.  
For the 3D case, the rupture tip position is computed by defining an equivalent length $L=\sqrt{A/\pi}$ as proportional to the square root of the rupture area $A$. 

The consequence is that ruptures starting from a larger nucleation area will accelerate more slowly, a feature that should be identifiable from far-field motion by retrieving the initial moment rate function. 

Further I posit that large nucleation can only take place on large faults, and offer arguments that the nucleation size should scale with the dimension of the fault segment where it takes place, at least, in the statistical sense. To this end I conduct a simple dimensional analysis of fracture energy on a fault with a self-affine distribution of relevant parameters, combined in a single dimensionless variable $\zeta$ (where $\zeta$ is proportional to the square of the stress drop and to the length of the crack). For the example model provided, the dimension of the initial rupture area does indeed scale linearly with the maximum size of the host fault.  

If the nucleation dimension is in some way controlled by the boundary conditions or the fault segmentation, then the initial nucleation should scale with the magnitude of the earthquake expected on such fault segment, to the extent that rupture stops a the fault segment border. This is not expected in a pure cascade model of rupture, but would apply in a case where the rupture process is somehow reflected by the preslip model. Inter-plate earthquakes appear to be more prone to preslip, because they are often preceded by protracted slow slip or activity of low intensity, and, therefore inter-plate earthquake may be more suitable candidates for the scaling derived here.  

Finally, given the above remarks, a slower initiation of rupture, which may be detected in the early seconds of far-field recordings, may be indicative of an increased probability of a large magnitude earthquake. Provided that the statistical significance of this scaling is verified by observation on catalogues of natural earthquakes, it would provide an interesting constraint on the early warning probabilistic forecast. 

\subsection*{Acknowledgements}I would like to thank Andrew Valentine for proof reading and helpful conversations on the limitations of probability estimates derived from small datasets, and Jeroen van Hunen for proof reading. I acknowledge support from Durham University for funding of open access publication. 

\printbibliography

\pagebreak
\section*{SUPPLEMENTARY INFORMATION}
\appendix

\section{Proof of time solution (\protect{\ref{eq:L_sol}})}
The derivative of the first branch of Lambert function $W_0$ is
\begin{equation}
\frac{\partial W_0(x)}{\partial t} = \frac{W_0(x)}{x\ \left(1 + W_0(x)\right)}
\end{equation} 
therefore the time derivative of $W_0\left[f(t)\right]$ is 
\begin{equation}
\frac{\partial W_0[f(t)]}{\partial t} = \frac{W_0\left[f(t)\right]\ f'(t)}{f(t)\ \left(1 + W_0\left[f(t)\right]\right)}
\end{equation} 
and taking the time derivative of solution (\ref{eq:L_sol}) we obtain:
\begin{equation}
\frac{\partial L}{\partial t} =C_{lim}\,\frac{W_0\left[e^{\frac{C_{lim}}{L_c}(t - t_0)}\right]}{W_0\left[e^{\frac{C_{lim}}{L_c}(t - t_0)}\right]+1}.
\label{eq:derofsol}
\end{equation}
We also have the original differential equation (\ref{eq:Ldiffeq}) into which we can replace $L$ by solution (\ref{eq:L_sol}) to obtain:
\begin{equation}
	{\setstretch{2}
%yn'[t] == Clim/yc (1 - 1/yn[t])
\frac{\partial L}{\partial t} = 
C_{lim}\ \left(1-\frac{L_c}{L}\right) = C_{lim}\ \left(1-\frac{1}{\left(W_0\left[e^{\frac{C_{lim}}{L_c}(t - t_0)}\right]+1\right)}\right).
\label{eq:Ldiffeq2}
}
\end{equation}
Equating (\ref{eq:derofsol}) to (\ref{eq:Ldiffeq2}) yields an identity, Q.E.D.
\section{Moment rate expression}
Starting from the expressin of the moment
\begin{equation}
M_o = C_a\ \mu\ \Delta U L(t)^2 = C_b\ \mu \frac{\Delta\tau\ L(t)}{\mu} L(t)^2 = C\Delta\tau\ L(t)^3
\end{equation}
(where $C_a, C_b$ and $C$ are geometrical factors) and taking the time derivative under the assumption that $\Delta\tau=\textrm{Const}$ and $C=\textrm{Const}$ yields
\begin{equation}
\dot{M}_o = 3 C \Delta\tau\ L^2(t)\ \dot{L}(t).
\end{equation}
Upon replacement of $L(t)$ by expression \ref{eq:L_sol} and $\dot{L}(t)$ by \ref{eq:Ldiffeq2} one obtains
\begin{equation}
\dot{M}_o = 3C\  C_{lim}\ \Delta\tau\ L_c^2 \left( W_o[e^{\frac{C_{lim}}{L_c}(t - t_0)}] ^2 + W_o[e^{\frac{C_{lim}}{L_c}(t - t_0)}] \right)
\end{equation}

\section{Experimental measurements of rupture velocitys}

I used Table 1 from Supplementary Material of  \citep{Latour2013}, reporting 
rupture length at the end of the quasi-static phase ($L_o$ in fourth column, that I equate to $L_c$ in my own derivations), the local rupture velocity $V_{loc}$ estimated between two neighboring sensors along the fault (fifth column), and the normal stress $\sigma_o$ (second column). 

Only the data where local rupture velocity $V_{loc}$ was reported in Table 1 of  \citep{Latour2013} was used. 
For $L_c$ I used the value of Table 1 when reported, and, when not reported, I computed an estimate $L_{c est}$ under the assumption that  
$L_c$ is inversely proportional to normal stress $\sigma_o$ such that $L_{c est} = P / \sigma_o$ with an arbitrary value $P=5\ 10^4$.

According to the analytical solution (\ref{eq:Ldiffeq}, main article), 
rewritten as $V_{loc} = C_{lim} + \frac{C_{lim}}{L} \ L_c$,
I expect a linear dependence between $V_{loc}$ and $L_c$ for fixed $C_{lim}$ and $L$. This is indeed apparent in Figure (\ref{fig:Latour_Vr_Lc}). 
The best fit line is obtained by adjusting the parameters $C_{lim}$ (velocity of shear waves in the polycarbonate of the experiment) 
and $L$ (the relative position of nucleation and sensors) that are not precisely known. The nucleation was always starting at the same location in the experiments of  \citep{Latour2013}, so we can assume that $L$ is approximately constant at the position of the piezolectric sensor.  

\section{Derivation of the approximate expression for $G(\gamma)$ \label{derivation}}

To the purpose of investigating scaling, we use the following definitions involving the static and dynamic stress intensity factors $K_0$ and $K$, and the static and dynamic energy flows $G_0$ and $G$:
\begin{equation}
{\setstretch{2.0}
	\begin{array}{lcr}
	G_0    & = & A_0\ \frac{K_0^2}{2\mu^\prime}\\      
	K_0    & = & \Delta\tau \sqrt{\pi\ L} 
	\end{array}
	\label{eq:intensity_energy_static}}
\end{equation}
\begin{equation}
{\setstretch{2.0}
	\begin{array}{lcr}
	G(v_r) & = & \frac{K^2}{2\mu^\prime}\, g_1(v_r)\\    
	K(v_r) & = & K_0 \,g_2(v_r)
	\end{array}
	\label{eq:intensity_energy_dynamic}}
\end{equation}
where $\Delta\tau$ is the stress drop inside the crack, $\mu'$ the shear stiffness and constant $A_0$ is a dimensionless shape factor involving loading geometry and Poisson ratio. Strictly speaking $g_1, g_2$ are functions of $v_r$, P, S and Rayleigh wave velocities; however for the 2D sub-sonic cases the velocity dependence can be simplified to $\gamma=v_r/C_{lim}$, the dimensionless ratio of rupture velocity to the limiting velocity $C_{lim}$ in the specific loading geometry, e.g., $\gamma=v_r/c_{Ral}$ for Mode II and $\gamma=v_r/c_S$ for Mode III (where $c_{Ral}$ and $c_S$ are the Rayleigh- and the shear-wave velocities, respectively. Expressing the velocity dependence as $w=g_1\ g_2^2$ we may write 
\begin{equation}
{\setstretch{1.5}
	\begin{array}{ccl}
	G(v_r) & = & w(\gamma)\ G_0
	\end{array}}
\end{equation}

Expressions of the energy flow for self-similar ruptures expanding at constant velocity have been derived in 2D and for cracks \citep{Kostrov1964, Broberg1999}, for self-healing pulses \citep{Nielsen2003} and in 3D for elliptical cracks \citep{Burridge1969}. 

More appropriately for this study, solutions of unstable accelerating ruptures for all three modes (I, II, III) in 2D are reported in \citep{Freund1990}, under the assumption that interaction originated from diffracted waves at the opposite rupture tips is negligible. 

\begin{figure}[h]
	\centering
	\includegraphics[width=0.7\linewidth]{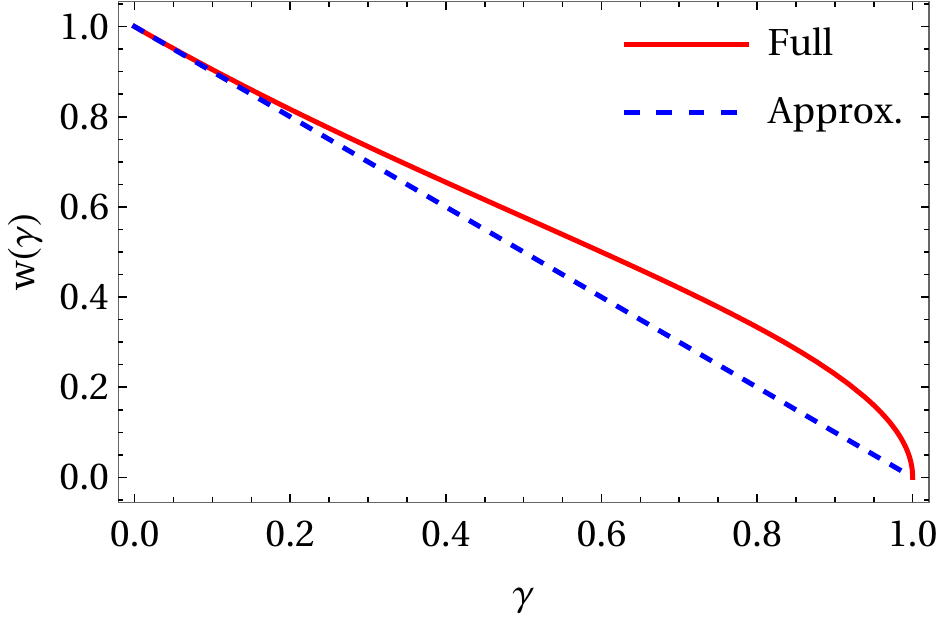}
	\caption{
	}
	\label{fig:approx_gvr}
\end{figure}

An analytical expression for the velocity dependence $w(\gamma)$ of energy flow $G= G_0\ w(\gamma)$ in mode III rupture can be obtained from equations (5.3.11) and (6.4.45) in  \citep{Freund1990}, resulting in 
\begin{equation}
w_\textrm{full}(\gamma)= \frac{\sqrt{1-\gamma}}{\sqrt{1+\gamma}}
\end{equation}
where $\gamma=v_r/C_{lim}$ and  in this specific geometry $C_{lim}$ is the shear wave velocity. Using the first order approximation in the vicinity of $\gamma=0$ one obtains the linear dependence:
\begin{equation}
w(\gamma)= {1-\gamma}
\label{eq:approx_gvr2}
\end{equation}
which is illustrated aginst the complete expression $w_\textrm{full}$ in Fig. (\ref{fig:approx_gvr}).

The in-plane (mode II) and tesile (mode I) geometries can also be approximated for all practical purposes by the linear trend of equation (\ref{eq:approx_gvr2}) 
in the interval $0<\gamma<1$ in the subsonic regime, where $C_{lim}$ is the Rayleigh wave velocity. 
The approximation can be obtained by taking the low velocity limit of equations (6.4.42, 5.3.311)  in  \citep{Freund1990} for mode II and similarly for  mode I (tension crack) according to equation (7.4.5), Ibid.

In the 3D case of spontaneously expanding shear crack, mixed mode II ans III are present. 
The sub-sonic crack develops with an anisotropic velocity resulting in a pseudo-elliptical rupture front  \citep{Favreau2002}.
Although no analytical solution is availalble for the energy flow in 3D (e.g. a solution exists for the self-similar elliptical crack under constant rupture velocity, see \cite{Burridge1969}), the stress intensity factor will scale with a characteristic length of rupture both in 2D and 3D (rupture radius for a circle or some radius-equivalent for a more complex shape). For 3D cracks, the stress intensity will combine any of the three factors ($K_I, K_{II}, K_{III}$) for individual modes, and expressions have been derived that use the minor semi-axis as the characteristic length and a corrective factor to account for the ellipse aspect-ratio, in the case of an elliptical crack \citep{Jwo2005}. 

To provide a suitable approximate solution in 3D, I use the same linear trend as in equation (\ref{eq:approx_gvr2}) and the geometrical mean of shear and Rayleigh wave velocities $C_{lim}=\sqrt{C_s\ C_{Ral}}$  in $\gamma=v_r/C_{lim}$ and define an equivalent radius, or average front position, as $L=\sqrt{A/\pi}$ where $A$ is rupture patch area.
I obtain an excellent fit to the numerical simulation by (see main paper).

In conclusion, it can be shown for all geometries that $w(\gamma)$ decreases monotonically from 1 to zero between $\gamma=0$ and $\gamma=1$, and that 
\begin{equation}
G(\gamma) \approx (1-\gamma)\ G_0
\label{eq:approx_flowdyn}
\end{equation}
is arguably a suitable approximation for the whole range of velocities, in particular for the initial acceleration of the breakout phase (see Appendix E). Expression (\ref{eq:approx_flowdyn}) has the desirable properties of monotonously decreasing with $v_r$ and exactly matching the full velocity function in the vicinity of $v_r=0$ for mode III, or within a constant close to 1 for mode II.
As a consequence a practical and general approximation for the crack tip motion equation can be rewritten as
\begin{equation}
\Gamma \approx (1-\gamma)\ G_0.
\label{eq:approx_flowdyn2}
\end{equation}
(an expression compatible the approximation proposed in \cite{Kammer2024} for the case where $C_{lim}=c_{Ral}$).

Note that I validate the approximation (\ref{eq:approx_flowdyn2}) by comparing it to rupture velocity obtained from a full 3D numerical simulation of spontaneous rupture. I also demonstrate numerically that the scaling in $L_c$ can be generalised to inhomogeneous parameter distributions on the fault and also applies non-trivial shapes of the initial breakout patch in 3D, where I define an equivalent radius, or average front position, as $L=\sqrt{A/\pi}$ where $A$ is rupture patch area.

\section{Self-affine $\chi$ and sub-fault areas}

\begin{figure}[h]
	\centering
	\includegraphics[width=0.7\linewidth]{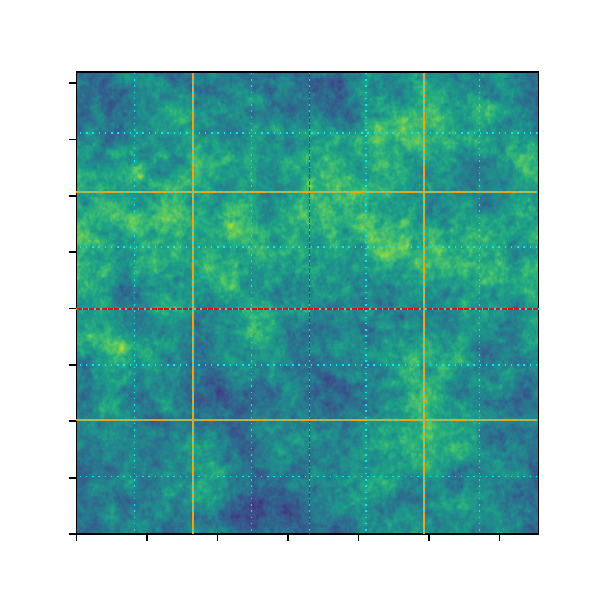}
	\caption{
		Illustration of sub-fault areas for the estimate of maximum $\zeta$.
		The color indicates the inhomogeneous value of $\chi$ on the fault surface.
		The vertical and horizontal lines show the subdivision of the fault in sub-faults
		of different sizes (dotted red: 4 sub-faults; yellow: 12 sub-faults; dotted blue: 64 sub-faults; further subdivisions not shown). The value of $\zeta$ is estimated 
		for each of the sub-faults in all subdivisions, and the dimension of the sub-fault with the maximum $\zeta$ is saved after each random realisation. The process is repeated 6 times
		for each of 5 upper dimension $L_f$ of the array ($L_f=$1024, 2048, 4086, 8192, 16384).
	}
	\label{fig:distri}
\end{figure}

\end{document}